\documentclass[journal]{IEEEtran}
\usepackage{cite}

\ifCLASSINFOpdf
   \usepackage[pdftex]{graphicx}
\else
\fi
\usepackage[cmex10]{amsmath}
\usepackage[tight,footnotesize]{subfigure}
\begin{document}
%
\title{Testing Hadronic Interaction Models using a \\ Highly Granular Silicon-Tungsten Calorimeter}
%
%
%

\author{Naomi van der Kolk \\ for the CALICE Collaboration
\thanks{N. van der Kolk is with the Max-Planck-Institute for Physics, Munich, Germany (telephone: +49 89 32354557, e-mail: kolk@mpp.mpg.de)}}

\maketitle

\begin{abstract}
\boldmath
The CALICE collaboration has published a detailed study of hadronic interactions using data recorded with the highly granular CALICE silicon-tungsten electromagnetic calorimeter (Si-W ECAL). Approximately 350,000 selected $\pi^{-}$ events at energies between 2 and 10\,GeV have been studied. The predictions of several physics models available within the {\sc Geant4} simulation tool kit are compared to this data. A reasonable overall description of the data is observed; the Monte Carlo predictions are within 20\,\% of the data, and for many observables much closer. The largest quantitative discrepancies are found in the longitudinal and transverse distributions of the reconstructed energy. Based on the good control of the data set and general observables, the next step is to achieve a deeper understanding of hadronic interactions by studying the interaction zone and by reconstructing secondaries that emerge from the hadronic interaction in the Si-W ECAL.
\end{abstract}


%

\section{Introduction}
%
%
%
%

\IEEEPARstart{T}{he} physics at future high-energy lepton colliders requires jet energy reconstruction with unprecedented precision. Detector concepts for the International Linear Collider (ILC) and the Compact Linear Collider (CLIC) rely on Particle Flow Algorithms \cite{Brient:2002gh,Thomson:2009rp} to achieve the necessary precision. This event reconstruction technique requires highly granular calorimeters to deliver optimal performance. Such calorimeters are developed and studied by the CALICE collaboration. 

To develop realistic Particle Flow Algorithms, the interactions of hadrons must be modelled reliably in Monte Carlo simulations and the detector response to hadrons must be well-understood. 
Highly granular calorimeter prototypes provide unique means to test and to further develop models of hadronic cascades. 

The response of the CALICE silicon-tungsten electromagnetic calorimeter prototype (Si-W ECAL)~\cite{Anduze:2008hq} is used to test hadronic shower models at low energies. The depth of the Si-W ECAL corresponds to approximately one interaction length ($\lambda_\mathrm{I}$), which means that, although the complete shower is not recorded, the first hadronic interaction can be studied in great detail because of the fine longitudinal and transversal sampling. 
The Si-W ECAL was operated in a test beam at Fermi National Accelerator Laboratory (FNAL) with negatively charged pions ($\pi^-$) in the energy of range 2 -- 10\,GeV. 
The majority of charged pions and other hadrons within high energy jets have energies in this range and therefore it is of considerable interest to validate the performance of Monte Carlo simulations.

\section{The CALICE Silicon-Tungsten Electromagnetic Calorimeter}

\begin{figure}[h]
  {\centering
    \includegraphics[width=0.4\textwidth]{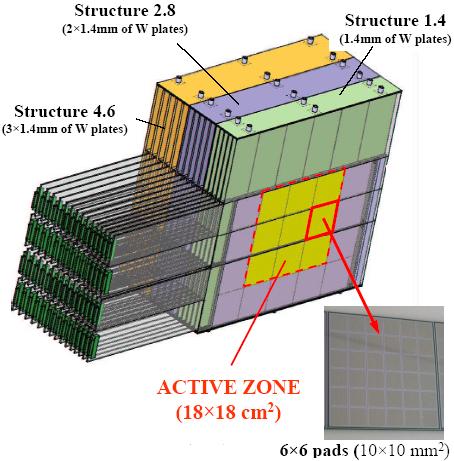}
    \caption{\sl Illustration of the Si-W ECAL physics prototype.}
    \label{figure:ecal}
  }
\end{figure}

The Si-W ECAL prototype, depicted in Fig.~\ref{figure:ecal}, consists of a sandwich structure of 30 layers of silicon as active material, alternating with tungsten as the absorber material.
The active layers are made of silicon wafers segmented into 1 $\times$ 1 cm$^{2}$ pixels (or pads). 
Each wafer consists of a square of 6 $\times$ 6 pixels and each layer contains a 3 $\times$ 3 matrix of these wafers, 
resulting in an active zone of 18 $\times$ 18 cm$^2$.

The Si-W ECAL is divided into three modules of ten layers each. 
The tungsten thickness per layer is different in each module, increasing from 1.4 mm in the first module (layers 1--10), 
to 2.8 mm in the second (layers 11--20) and 4.2 mm in the third (layers 21--30). 
The total thickness corresponds to 24 radiation lengths (X$_{0}$) and approximately one interaction length. 
More than half of the hadrons traversing the Si-W ECAL prototype undergo a primary interaction within its volume.

\section{Results}
Test beams were conducted in May and July of 2008 at the Fermilab Test Beam Facility at FNAL. 
The presented analysis, detailed in the CALICE publication~\cite{Bilki:2014uep}, uses data from runs with $\pi^{-}$ mesons at energies of 2, 4, 6, 8 and 10\,GeV. 
Monte Carlo simulations corresponding to the recorded test beam data have been produced using the simulation tool kit {\sc Geant4}~\cite{Geant:2003} (version 9.6 patch 1).
The full geometry of the CALICE test beam set-up is taken into account in the simulation via the {\sc mokka} framework which provides the geometry interface to {\sc Geant4}. 
Four different physics lists,  {\sc ftfp\_bert}, {\sc ftfp\_bert\_hp}, {\sc qgsp\_bert} and {\sc qbbc}, are compared to the data in order to study the different hadronic models which are implemented in the physics lists for different energy ranges. 

As a first step in the analysis a pure data sample is selected. Residual contamination from electron events and events with multiple incoming particles is corrected for.
Subsequently interacting events are selected based on the deposited energy in individual layers of the Si-W ECAL. 
Two selection criteria are applied: one based on the increase of the absolute energy in subsequent layers, and one based on the relative increase in deposited energy.
First a requirement is made on the reconstructed energy in each layer, $E_{i}$.
If three consecutive layers have an energy higher than a threshold, $E_{cut}$, the interaction layer is identified as the first of these (layer $i$).
This selection is not efficient at lower beam energies, where a small number of low energy secondaries is produced and shower fluctuations are relatively strong.
A second selection criterion based on the relative increase in reconstructed energy is applied to events not selected by the first criterion.
The efficiency to define interacting events improves considerably at the lower energies studied when this second criterion is applied; for 2\,GeV the efficiency increases from 35\% to 60\%.
The efficiency increases with beam energy from 60\% at 2\,GeV to 93\% at 10\,GeV.
The starting layer of the hadronic shower can be reconstructed with an accuracy of $\pm$ 2 layers at an efficiency of at least 50\% at 2\,GeV and 87\% at 10\,GeV.

The same event selection criteria as used for the data are also applied to the simulated events. The predictions from the simulations are then compared to the data.
The fraction of interacting events and the total deposited energy are studied, as well as radial and longitudinal shower profiles and hit distributions.

 The depth of the Si-W ECAL corresponds to approximately one interaction length ($\lambda_\mathrm{I}$), which means that more than half of the pions traversing the prototype will have an interaction. This fraction of interacting events is measured to be independent of energy and is reproduced by the simulations for low energies while for higher energies the simulations overestimate the fraction by 7\%.
 
\begin{figure}[h]
  {\centering
    \includegraphics[width=0.48\textwidth]{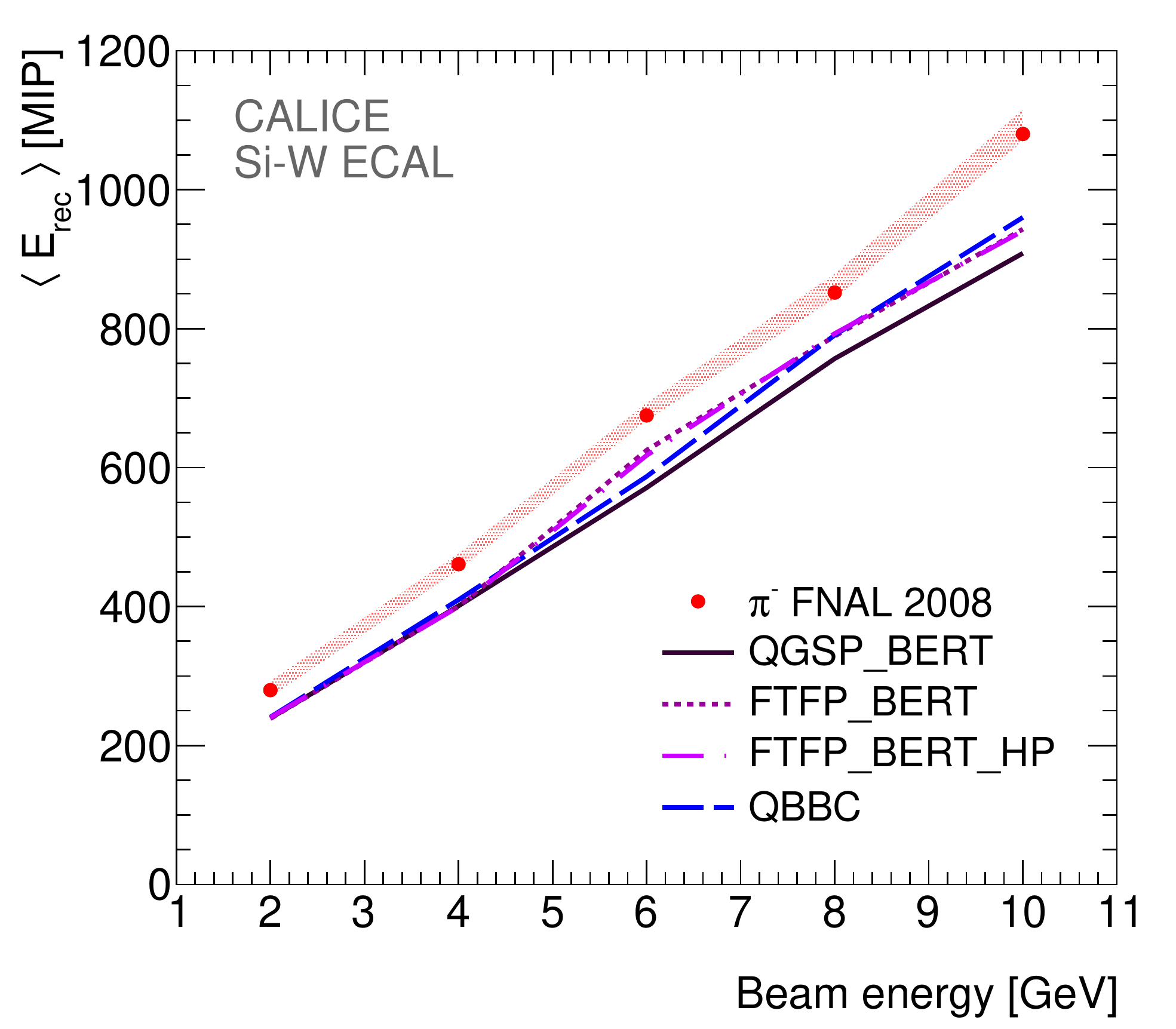}
    \caption{\sl Reconstructed $\pi^-$ shower energy in the Si-W ECAL for data and various Monte Carlo physics lists as a function of beam energy (2\,GeV to 10\,GeV).}
    \label{figure:showerenergy}
  }
\end{figure}

 The energy deposited by the pion showers is underestimated by the simulations.
 Fig.~\ref{figure:showerenergy} shows the reconstructed energy of that part of the shower that is seen in the Si-W ECAL as a function of beam energy. 
 This shower energy increases with the energy of the incoming $\pi^-$ and is on average 15\% higher in data than in simulation. 

\begin{figure}[h]
  {\centering
    \includegraphics[width=0.48\textwidth]{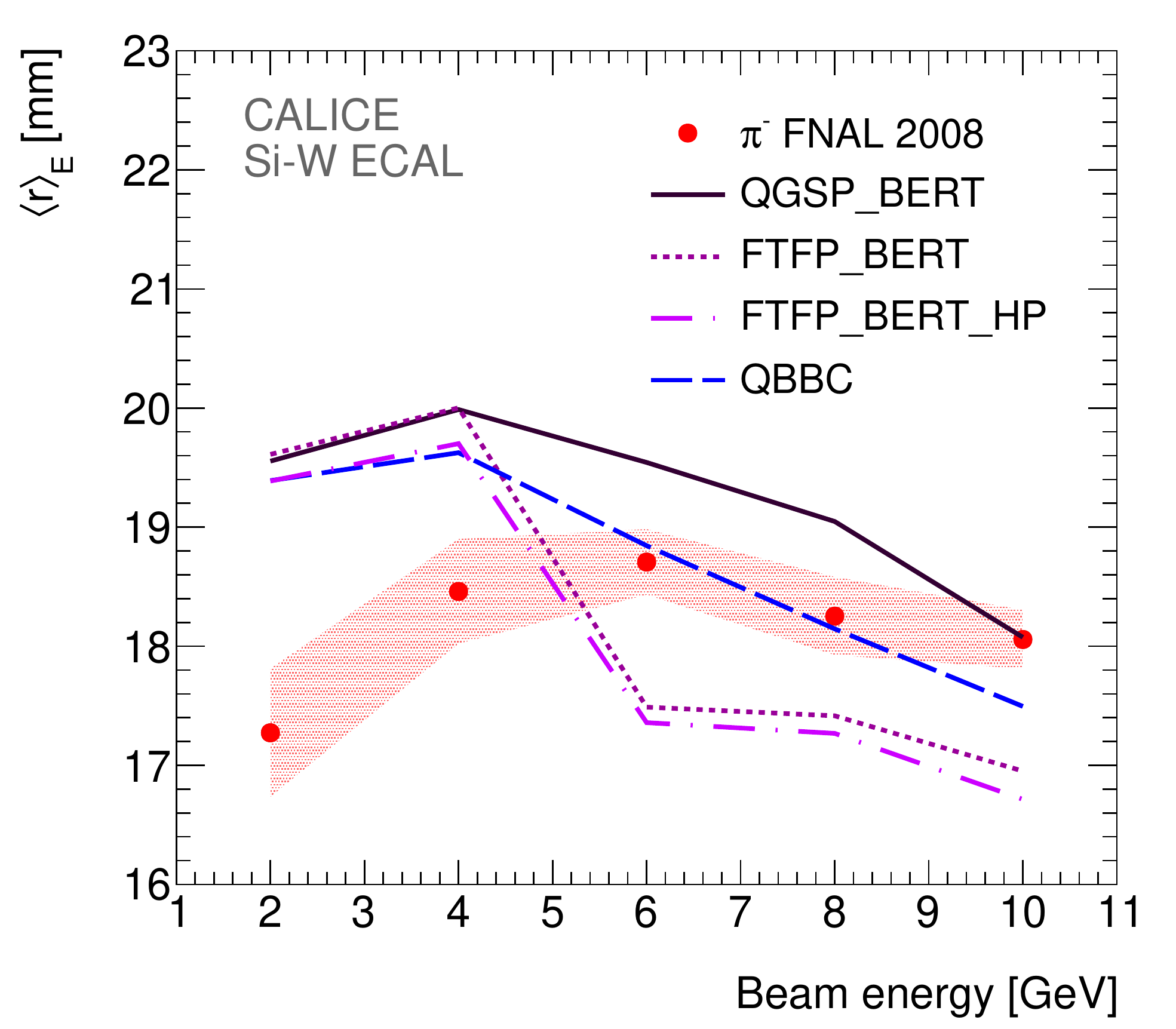}
    \caption{\sl Mean of the radial energy profile for interacting events as a function of beam energy (2\,GeV to 10\,GeV) for data and various Monte Carlo physics lists.}
    \label{figure:meansigmaradialprofile}
  }
\end{figure}

The radial hit and energy distributions prove to be sensitive to the different hadronic models implemented in the {\sc Geant4} physics lists, while the longitudinal distributions are not sensitive to these transitions.
Fig.~\ref{figure:meansigmaradialprofile} shows the mean of the radial energy profiles as a function of the beam energy.
The model transition between the Bertini cascade (up to 4\,GeV) and Fritiof string model in {\sc ftfp\_bert} and {\sc ftfp\_bert\_hp} between 4 and 6\,GeV is very pronounced.
The transition from the Bertini cascade to the Low Energy Parametrised model is {\sc qgsp\_bert} between 8 and 10\,GeV is also visible.
The same behaviour is seen in the mean of the radial hit distributions, which are described by the simulation within 6\% of the data.

\begin{figure}[h]
  {\centering
    \includegraphics[width=0.48\textwidth]{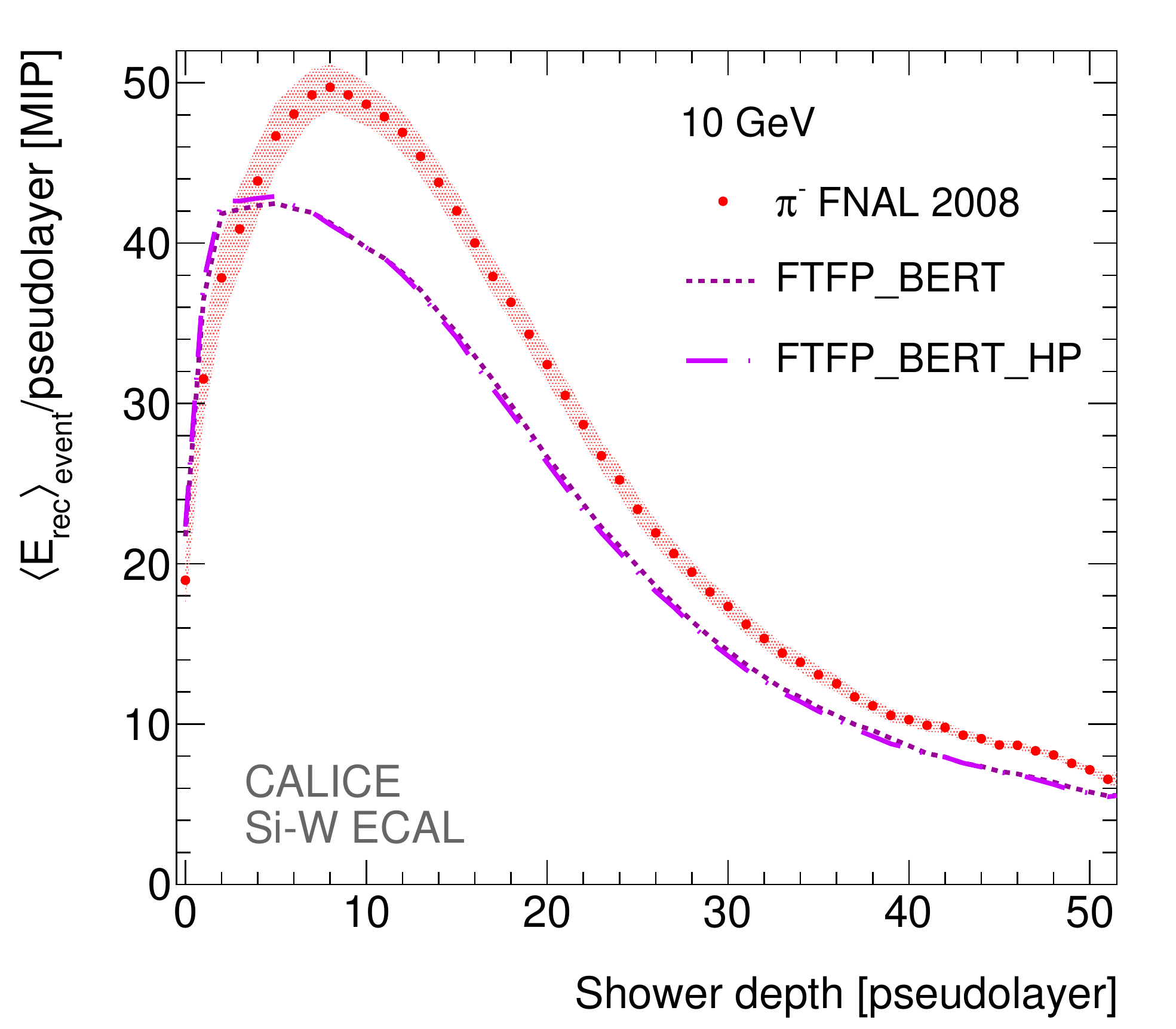}
    \caption{\sl The Longitudinal energy profile for interacting events at 10\,GeV for data and the Monte Carlo physics lists {\sc ftfp\_bert} and {\sc ftfp\_bert\_hp}.}
    \label{figure:longitudinalprofile10GeVFTFP}
  }
\end{figure}

\begin{figure}[h]
  {\centering
    \includegraphics[width=0.48\textwidth]{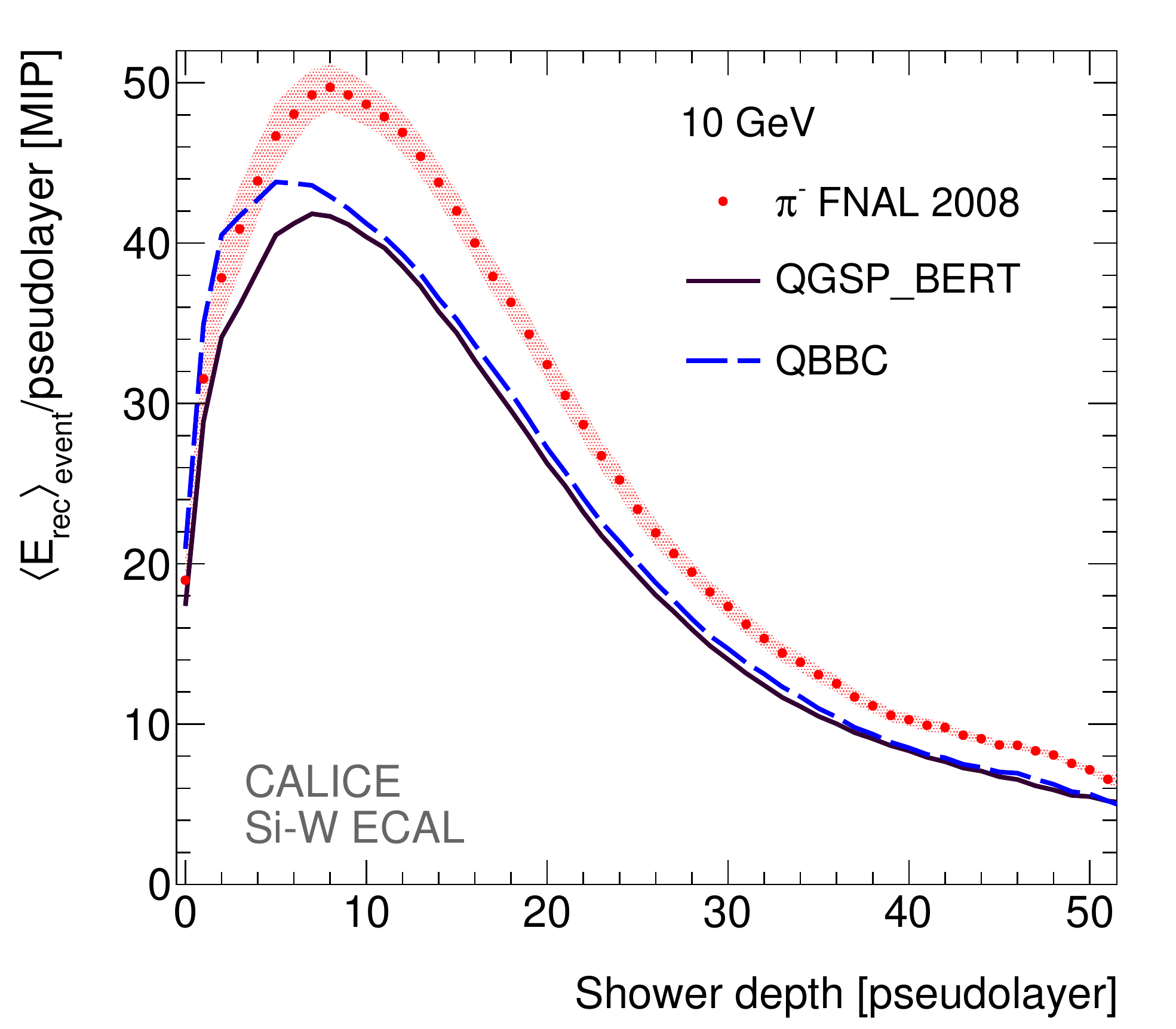}
    \caption{\sl The Longitudinal energy profile for interacting events at 10\,GeV for data and the Monte Carlo physics lists {\sc qgsp\_bert}  and {\sc qbbc}.}
    \label{figure:longitudinalprofile10GeVQGSP}
  }
\end{figure}

The shape of the hit distributions, longitudinally and radially, is well reproduced by simulations, only for the radial hit distributions the mean is overestimated.
In contrast, the energy profiles are not reproduced so well.
This is most pronounced in the longitudinal energy profile, where the deposited energy in simulations is up to 16\% too low.
This energy deficiency is seen in all the studied physics lists and is larger for higher energies.
Fig.~\ref{figure:longitudinalprofile10GeVFTFP} and Fig.~\ref{figure:longitudinalprofile10GeVQGSP} show the longitudinal energy profile at 10\,GeV for {\sc ftfp\_bert} and {\sc ftfp\_bert\_hp} compared to data and {\sc qgsp\_bert}  and {\sc qbbc} compared to data. 
While the energy deposition is in general too low, for {\sc ftfp\_bert}, {\sc ftfp\_bert\_hp} and {\sc qbbc} the energy deposition is too high in the first few layers after the interaction.

Especially for {\sc ftfp\_bert} the energy deficiency is unexpected, because in a previous version of {\sc Geant4} the energy was well reproduced for this prototype.
Between these two versions the Fritiof String Model was heavily modified and was tuned using thin target scintillator data. 
The observed discrepancy with the Si-W ECAL could be related to the sensitive material, which is silicon, as such a discrepancy is not seen in the CALICE scintillator-tungsten HCAL prototype~\cite{Adloff:2014}.

\begin{figure}[h]
  {\centering
    \includegraphics[width=0.48\textwidth]{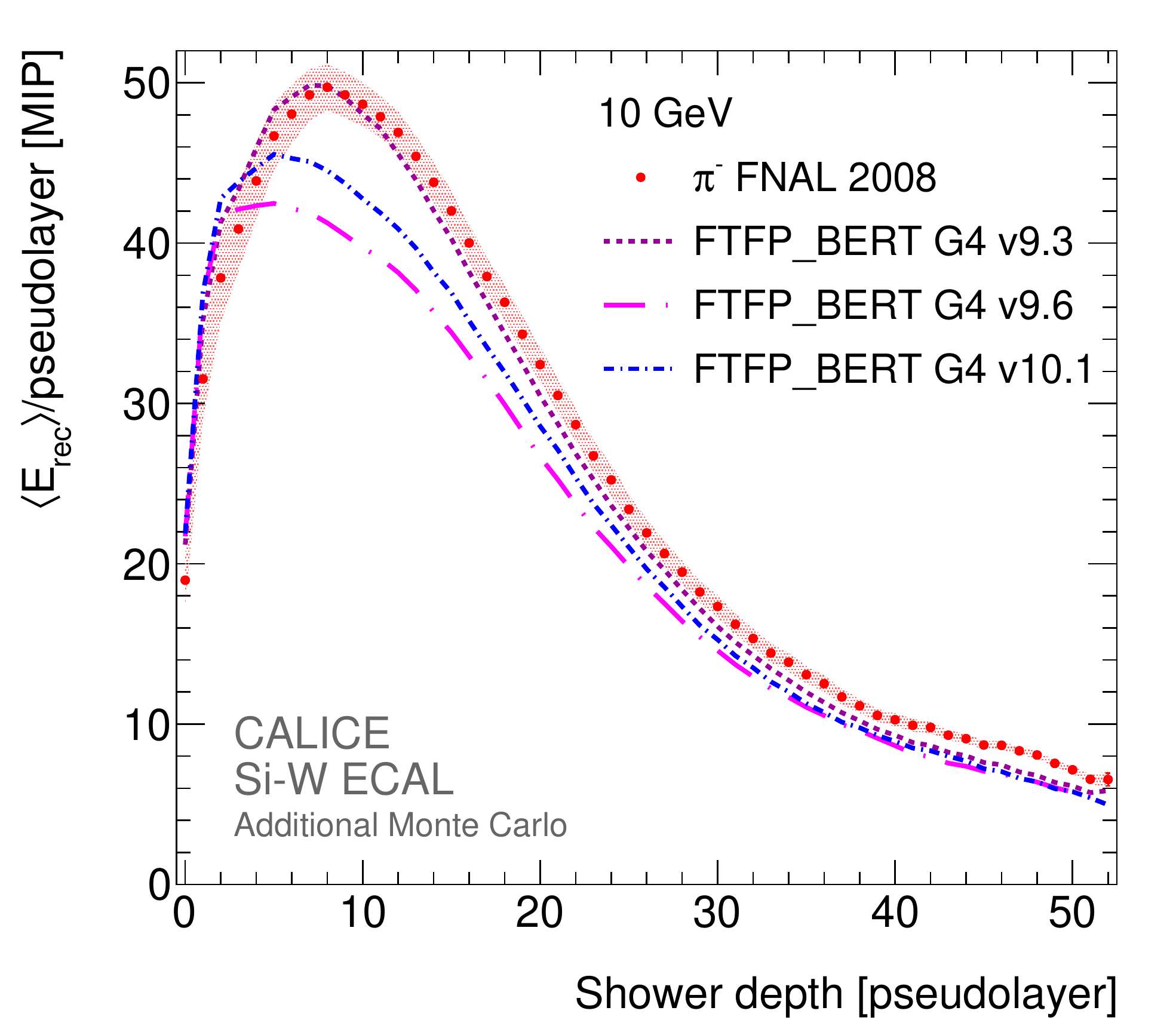}
    \caption{\sl The longitudinal energy profile for interacting events at 10\,GeV for data and the Monte Carlo physics list {\sc ftfp\_bert} for 3 versions of {\sc Geant4}.}
    \label{figure:longitudinalprofile10GeVcompared}
  }
\end{figure}

Very recently the newest version of {\sc Geant4} became available for comparison.
Fig.~\ref{figure:longitudinalprofile10GeVcompared} shows the longitudinal energy profile at 10\,GeV of data and {\sc ftfp\_bert} for 3 different versions of {\sc Geant4}; version 9.3, 9.6 and 10.1.
While version 9.3 describes the data best, version 10.1 is an improvement upon version 9.6, but the energy deposition is still too low.

\begin{figure}[h]
  {\centering
    \includegraphics[width=0.48\textwidth]{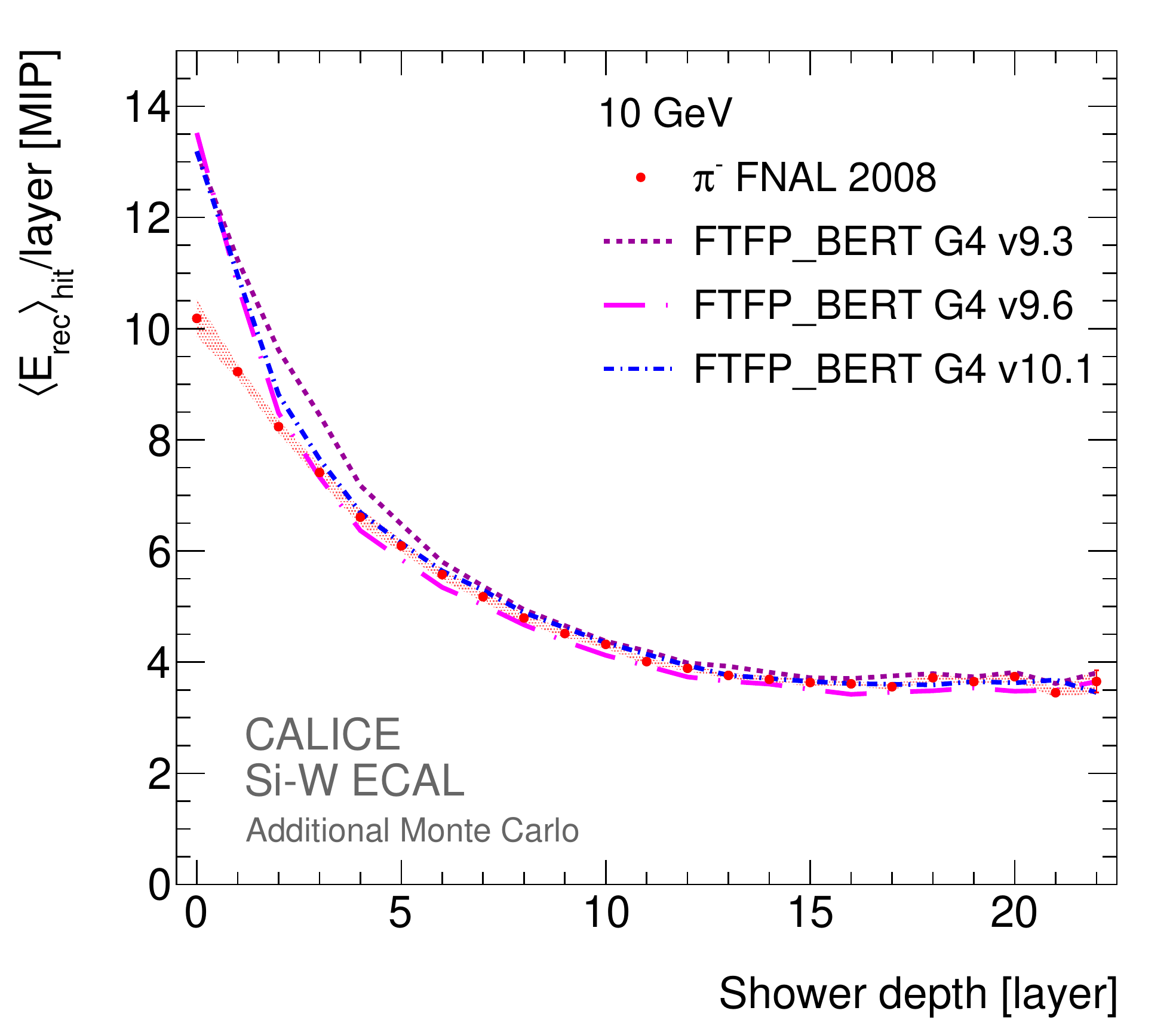}
    \caption{\sl The longitudinal mean hit energy for interacting events at 10\,GeV for data and the Monte Carlo physics list {\sc ftfp\_bert} for 3 versions of {\sc Geant4}.}
    \label{figure:longitudinalmeanhitenergy10GeVcompared}
  }
\end{figure}

The mean hit energy in the showers is very similar in data and simulations starting from the fifth layer after the interaction, as can be seen in Fig.~\ref{figure:longitudinalmeanhitenergy10GeVcompared}, closer to the interaction the energy per hit is too high in the simulations. 
Fig.~\ref{figure:longitudinalmeanhitenergy10GeVcompared} shows the mean hit energy per layer for data at 10\,GeV, compared to {\sc ftfp\_bert} for the same 3 versions of {\sc Geant4}.

For version 9.3 the hit energy is always above that in the data and the excess close to the shower start is larger.
For version 9.6 the hit energy is slightly lower than the data except for the first few layers, and version 10.1 describes the data best overall.
The too high hit energy can explain the energy excess in the longitudinal energy profile in the first few layers.
The similar mean hit energy for all other layers points to a deficit in the number of hits produced in the simulations compared to the data.

\begin{figure}[h]
  {\centering
    \includegraphics[width=0.48\textwidth]{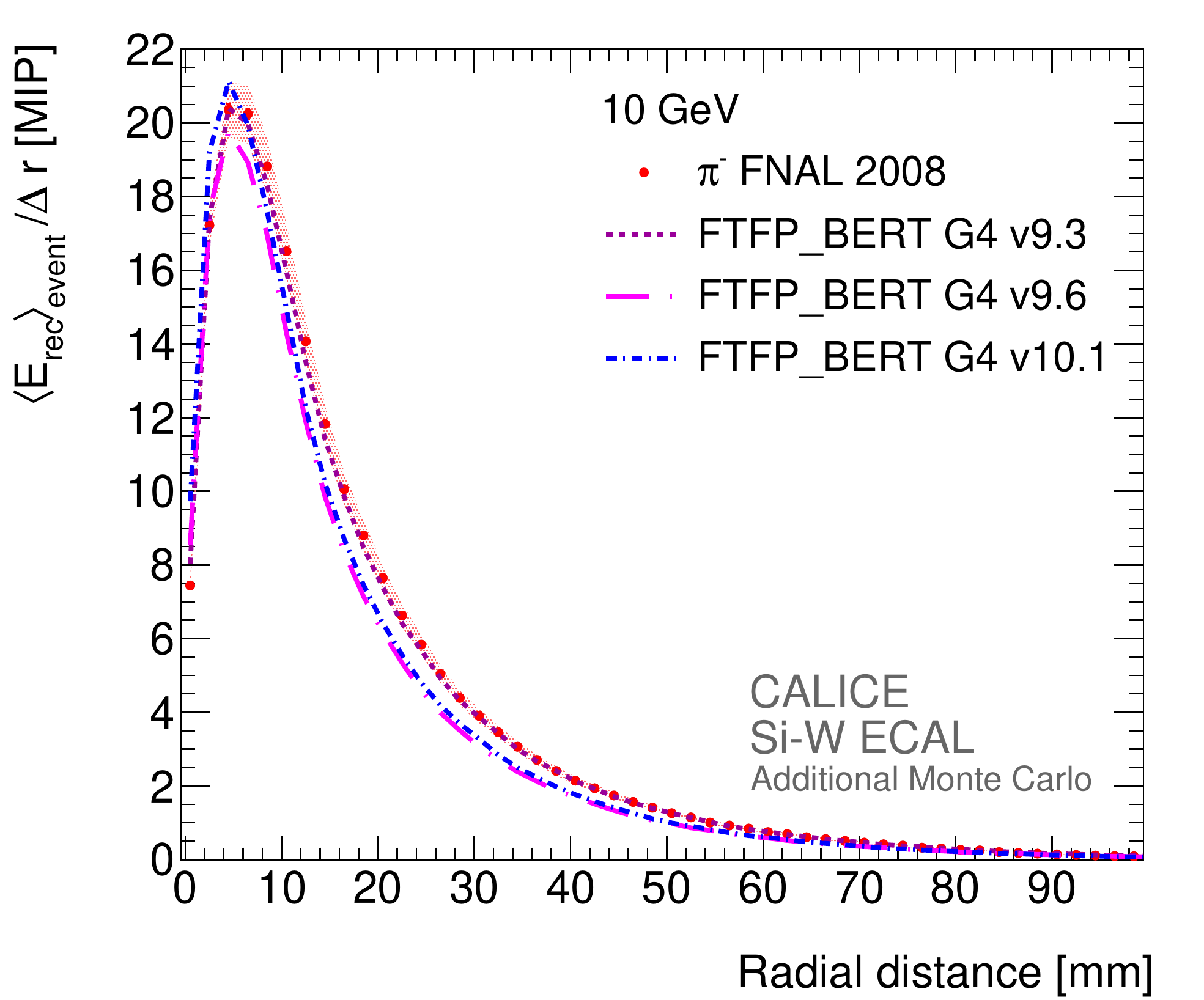}
    \caption{\sl The radial energy profile for interacting events at 10\,GeV for data and the Monte Carlo physics list {\sc ftfp\_bert} for 3 versions of {\sc Geant4}.}
    \label{figure:radialenergyprofile10GeVcompared}
  }
\end{figure}

Also in the radial energy profile the energy is underestimated in version 9.6 and 10.1, however not so much.
The radial energy profile at 10\,GeV is shown is Fig.~\ref{figure:radialenergyprofile10GeVcompared} for data and 3 versions of {\sc ftfp\_bert}.

\begin{figure}[h]
  {\centering
    \includegraphics[width=0.48\textwidth]{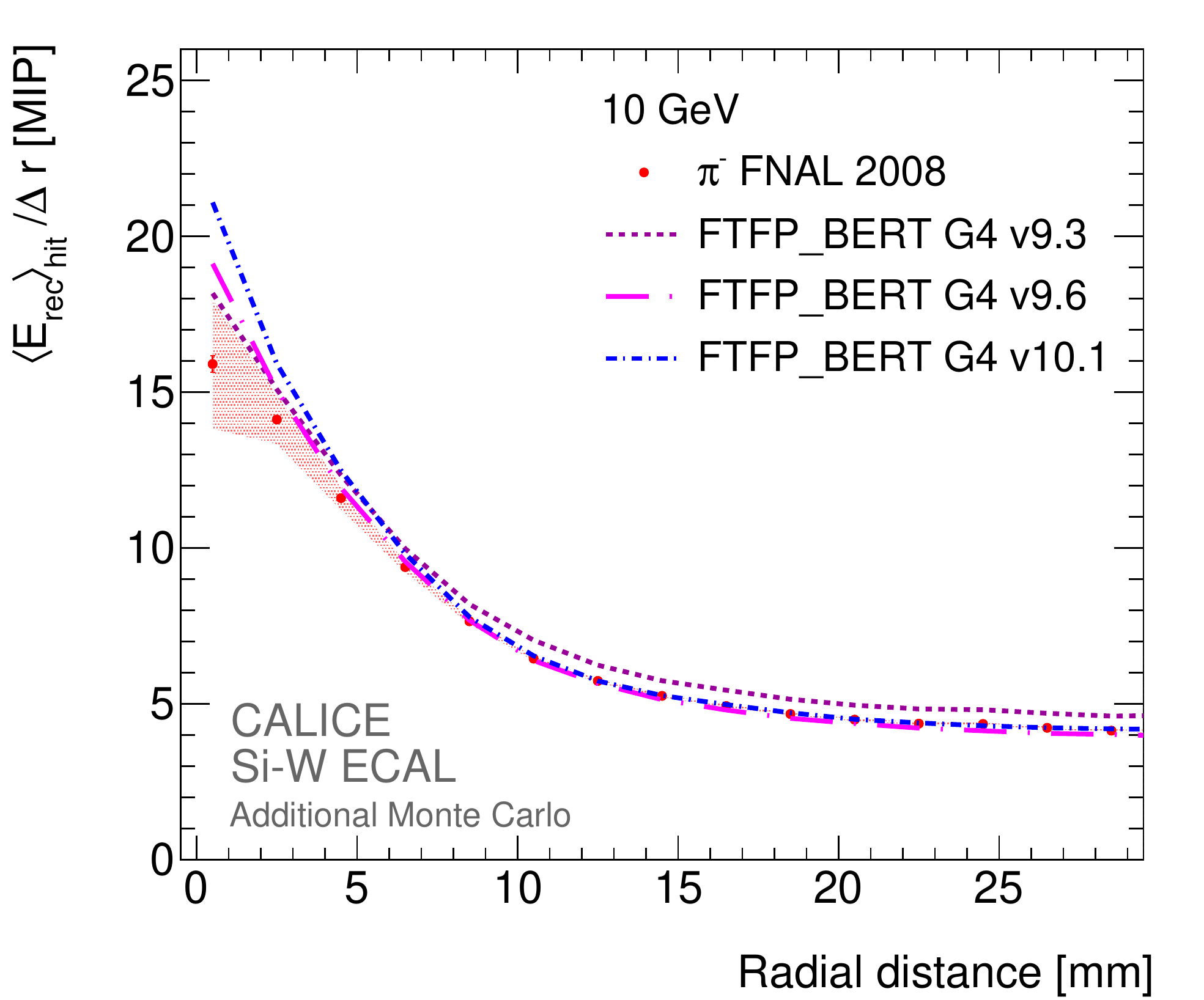}
    \caption{\sl The radial mean hit energy for interacting events at 10\,GeV for data and the Monte Carlo physics list {\sc ftfp\_bert} for 3 versions of {\sc Geant4}.}
    \label{figure:radialmeanhitenergy10GeVcompared}
  }
\end{figure}

Fig.~\ref{figure:radialmeanhitenergy10GeVcompared} shows the radial mean hit energy.
Close to the shower core (small radii) the mean hit energy is too high in the simulation.
For version 9.3 of {\sc Geant4} the overestimation is smallest while it is highest in version 10.1.
The tail of the distribution, on the other hand, is described better in version 9.6 and 10.1 than in 9.3 where the mean hit energy is too high.

\section{Studying the structure of the first hadronic interaction}
\begin{figure}[h]
  {\centering
    \includegraphics[width=0.48\textwidth]{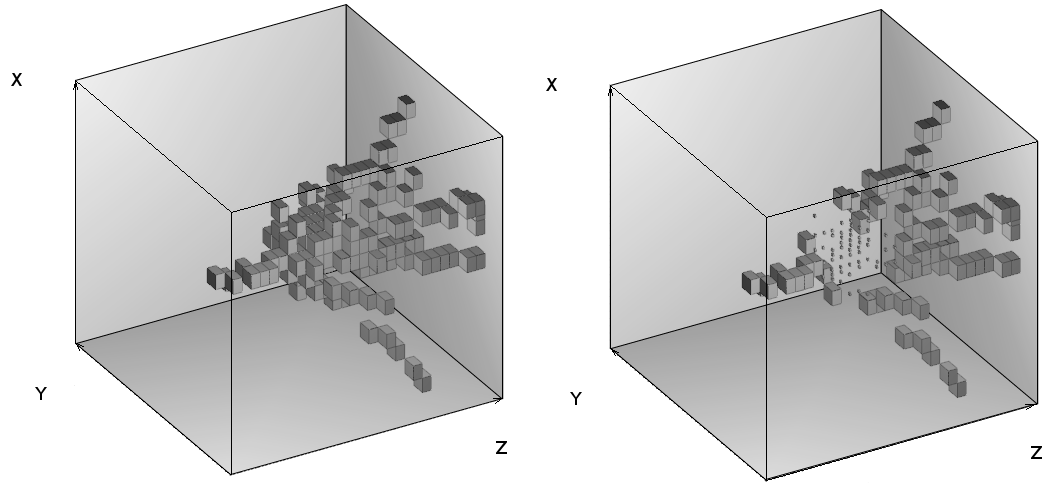}
    \caption{\sl Event display of the interaction of a primary pion with an energy of 10\,GeV in the Si-W ECAL. The interaction leads to a dense interaction zone (removed on the right hand side) and secondaries that can be reconstructed as outgoing tracks.}
    \label{figure:hadshow}
  }
\end{figure}

After the study of general observables of hadronic showers an even more detailed analysis is under way.
An algorithm has been developed to measure tracks in the Si-W ECAL while permitting at the same time to characterise the interaction zone in terms of energy and extension. 
Fig.~\ref{figure:hadshow} shows an event in the Si-W ECAL with an incoming 10\,GeV primary pion that interacts and produces secondaries. 
On the right hand side the interaction region has been defined and hits belonging to this region have been removed for better visualisation of the secondary tracks.

\begin{figure}[h]
  {\centering
    \includegraphics[width=0.48\textwidth]{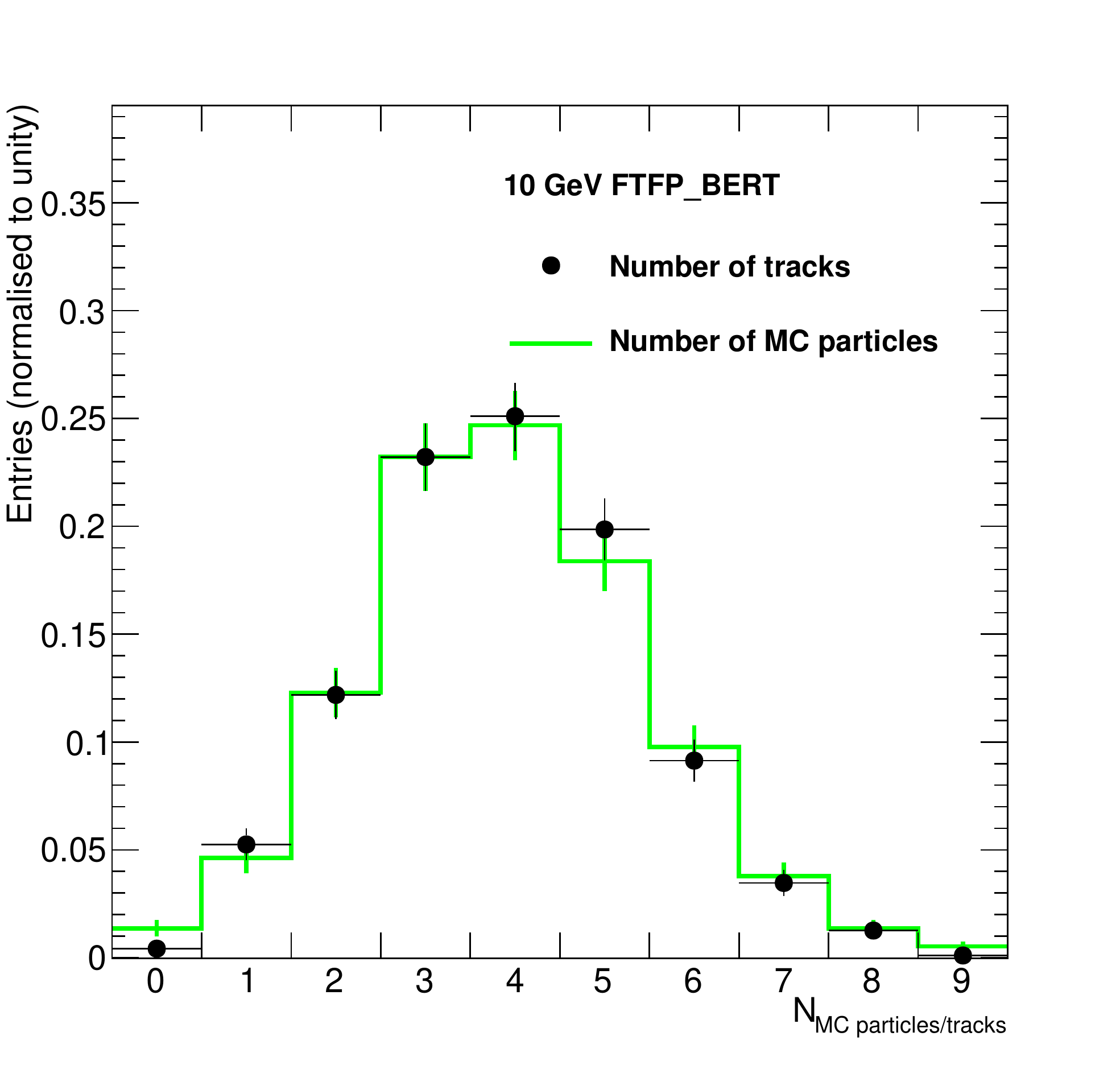}
    \caption{\sl The number of generated and reconstructed secondary tracks for 10\,GeV pions in {\sc ftfp\_bert}.}
    \label{figure:tracks}
  }
\end{figure}

The track finding algorithm has been tested using the generator level information from simulated events.
Fig.~\ref{figure:tracks} shows the number of generated and reconstructed tracks and shows that the generated tracks are properly identified by the track finding algorithm.
All observables studied in this analysis are very detailed probes of the hadronic shower models used in the simulation.  
For the first time the analysis will compare the energy deposited in the interaction zone and the number of tracks that emerge from the interaction zone for different energies on the primary pion.  
Preliminary results show that, as expected, the deposited energy and the number of tracks increase with the energy of the primary pion.

\section{Conclusion}

The CALICE Si-W ECAL prototype has been shown to deliver very detailed information on the interaction of pions.
Interactions in the energy range of 2 to 10\,GeV have been studied in terms of shower observables such as the deposited energy and radial and longitudinal distributions. 
These distributions are compared to simulations with several physics lists in {\sc Geant4}.
None of the studied physics lists describe the entire set of data, but overall the simulations are within 20\% of the data and for most observables much closer.
The longitudinal hit distribution is very well described, while the mean is shifted for the radial hit distribution.
On the other hand the physics observables that take into account the energy deposition are not reproduced well in simulations.
Based on the average energy deposited per hit, which agrees between data and simulation, the lower total energy in the simulations can be attributed to a lower number of hits.
By combining the longitudinal and radial energy profiles it seems that especially the Fritiof model, implemented in {\sc ftfp\_bert} and {\sc ftfp\_bert\_hp} above 4\,GeV, deposits too much energy very close to the interaction region. 
The radial distributions prove to be very sensitive to the transitions between different hadronic models in the physics lists.

In conclusion, the data from the Si-W ECAL are very precise and allow for discriminating between {\sc Geant4} models on a very fine scale.
This enables further analysis into hadronic showers that will characterise the interaction region in terms of energy and size and reconstruct the produced secondaries (tracking). 
Precise tracking in the Si-W ECAL is an indispensable premise for the application of Particle Flow algorithms to highly granular calorimeters. 
The successful seed of secondary tracks in the Si-W ECAL facilitates the reconstruction of tracks throughout the calorimeter system that will be completed by a hadronic calorimeter with a depth of several interaction lengths. 
The secondaries that lead to measurable tracks interact in the Si-W ECAL in form of MIPs. 
Thus the analysis is suited to serve as a first step towards an in-situ calibration of the Si-W ECAL, that could be used to monitor the response of a full size Si-W ECAL in a Linear Collider experiment.


%




\ifCLASSOPTIONcaptionsoff
  \newpage
\fi



%



\bibliographystyle{IEEEtran.bst}
\bibliography{CALICE}

%








\end{document}